\documentclass{article}
\usepackage{epsfig}
\begin{document}
\begin{flushright}
\normalsize
Freiburg-THEP 01/02\\
\end{flushright}

\begin{center}
{\Large\bf GRAVITATING SPHALERONS AND SPHALERON BLACK HOLES IN
ASYMPTOTICALLY ANTI-DE SITTER SPACETIME }
\vspace{1.2cm}\\
J.J.\ van der Bij and Eugen Radu\footnote{\textbf{corresponding author}:
\\Albert-Ludwigs-Universit\"at Freiburg, Fakult\"at f\"ur Physik, 
\\Hermann-Herder-Stra\ss e 3, D-79104 Freiburg, Germany
\\ email: radu@newton.physik.uni-freiburg.de
\\ Telephone: +49-761/203-7630 
\\Fax: +49-761/203-5967}\vspace{0.4cm}\\
\it Albert-Ludwigs-Universit\"at \\
\it Fakult\"at f\"ur Physik\\
\it Freiburg Germany\vspace{1.2cm}\\
\end{center}

\begin{abstract}
Numerical arguments are presented for the existence of spherically symmetric regular and black hole
solutions of the EYMH equations with a negative cosmological constant.
These solutions approach asymptotically the anti-de Sitter spacetime. The main properties 
of the solutions and the differences with respect to the asymptotically flat case are discussed.
The instability of the gravitating sphaleron solutions is also proven.
\\
\\
\end{abstract}
\textbf{PACS}: 04.40; 14.80.B, C; 98.80.E
\newpage 
\section{INTRODUCTION}

Investigation of classical solutions of the Yang-Mills theory coupled to 
gravity revealed many unexpected features both 
in soliton and black hole physics. 
After Bartnik and McKinnon discovered a nontrivial particlelike solution of 
the Einstein-Yang-Mills (EYM) equations \cite{bmk}, there has been a great deal of 
numerical and analytical 
work on various aspects of EYM theory and a variety of self-gravitating 
structures with non-Abelian fields have been found (for a review see \cite{1}).
These include black holes with non-trivial hair, thereby leading to the 
posibility of evading the no-hair conjecture.

Most of these investigations have been carried out on the assumption that spacetime 
is asymptotically flat.
Less is known when the theory is modified to include a cosmological constant 
$\Lambda$ which greatly changes the asymptotic structure 
of spacetime \cite{Hawking}.

For a positive cosmological constant, the behavior of cosmological solutions of 
EYM equations is similar in many 
respects to that for asymptotically flat geometries \cite{2} and in particular the
 configurations are also unstable \cite{3}.

Although in the last years has been a lot of interest in asymptotically anti-de Sitter
 (AdS) spacetimes, interest widely connected with the string theory and related topics,
the classical solutions of a nonabelian field theory in this background seems 
to be less studied.

Recently, some authors have discussed the properties of soliton  and 
black hole solutions of the EYM system for $\Lambda <0$, 
$i.e.$ an asymptotically AdS spacetime \cite{4, 5, 6}. 
They obtained some surprising results, which are rather different from the 
corresponding ones for asymptotically flat and de Sitter spacetimes.
First, there are solutions for continuous intervals of the parameter space, 
rather then discrete points. 
Secondly there are nontivial solutions stable against spherically symmetric 
linear perturbations corresponding 
to stable monopole and dyon configurations.

Given the different behavior of the EYM system in an asymptotically AdS 
spacetime it is natural to approach the study of this system in the presence 
of a Higgs field. 

As is well known, no regular particle-like solutions of the SU(2) Yang-Mills 
 equations exist in Minkowski spacetime \cite{7}.
Physically this can be understood as a consequence of the repulsive nature of 
the Yang-Mills vector field.
When scalar fields are added, regular solutions become possible due to the
 balance of the YM repulsive force and the attractive force of the scalars.
Two types of such solutions are known: magnetic monopoles and sphalerons.

In ref. \cite{8, 9} regular gravitating monopole and dyon solutions in 
Einstein-Yang-Mills-Higgs (EYMH) theory with Higgs field 
in the adjoint representation were shown to exist in asymptotically 
AdS spacetime. 
As it happens in asymptotically flat space, a critical value for the Newton 
constant exists above which 
no regular solution can be found.
The presence of a cosmological constant enhances this effect, the critical 
value being smaller than the asymptotically flat one. 

In complete analogy to gravitating monopoles one may consider self-gravitating
sphalerons corresponding to a Higgs field doublet.
In ref. \cite{10} strong numerical arguments were presented for the existence of both 
regular and black-hole solutions to Einstein gravity (without a cosmological term) 
coupled to SU(2) gauge theory and a Higgs doublet, as in the standard model.
For each fixed value of the Higgs vacuum expectation value $v$, 
solutions have been found, that can be indexed by the number of nodes $k$ of 
the Yang-Mills potential function.
For each $k$ there are two branches of solution, depending on the behavior of 
 $v\to 0$. 
The two branches of solutions converge for some values of the theory 
parameters \cite{1, 11}. 
This system has been also examined from the stability point of view and 
found to be unstable \cite{12, 13}. 

Because of the physical importance of these objects, it is worthwhile to study
generalizations in a different cosmological background.
We may expect that similar to the EYM theory, the different asymptotic structure 
of the spacetime will affect the properties of the solutions leading to some 
new effects. 

In this paper we study both analytically and numerically  regular and black-hole  
solutions of the coupled  EYMH field equations with a negative cosmological 
constant, extending the results of ref.\cite{10} to this case. 

Different from the asymptotically flat case, the EYMH and EYM systems seem to 
have solutions with rather different properties; the numerical solutions we have found do 
not retain the nontrivial properties of the pure EYM case.
In particular they have no YM charge and no electric part of the YM potential.

Although the results we shall find are broadly similar to those valid for the 
$\Lambda=0$ case, there are some differences.

A nonzero $\Lambda$  term in the action implies a complicated power decay of the 
fields at infinity, rather than exponentially as expected.
Also, the parameter range of the solutions found in ref.\cite{10} remains no 
longer valid and a new range has to be found for every choice of $\Lambda$. 
The existence of a nonzero cosmological constant implies a decrease of the 
maximal allowed vacuum  expectation value of the Higgs field.

The  paper is structured as follows:  in section 2 we present the general 
framework and analytically analyse the field equations, while in section 3 we address 
the problem of the numerical  construction  of  solutions.  
In section 4 the stability of the solutions is considered. 
The solutions are found to be unstable. 
We conclude with section 5 where the results are compiled.
 
\section{GENERAL FRAMEWORK AND BASIC EQUATIONS}

\subsection{Basic ansatz}
Our study of the EYMH system is based upon the action
\begin{equation} \label{lag0}
S=\int d^{4}x\sqrt{-g}[\frac{1}{16\pi G}(\mathcal{R} - 2 \Lambda)-
\frac{1}{4\pi}((D_{\mu}\Phi)^{\dagger}(D^{\mu}\Phi)+
V(\Phi))-\frac{1}{4\pi}\frac{1}{4} |F|^{2}]. 
\end{equation}
Here  $G$ is  the  gravitational constant,  $D_{\mu}$ is  the  usual  gauge-covariant
derivative expressed in the anti-hermitian basis of SU(2) ($\tau_{a}=-i\sigma_{a}/2$)
\begin{equation}
D_{\mu}=\partial_{\mu}+g\tau\cdot A_{\mu},
\end{equation}
$g$ is the gauge coupling constant. 

As we assume spherical symmetry it is convenient to use the usual metric form
\begin{equation} \label{metric}
ds^{2}=\frac{dr^{2}}{H(r)}+r^2(d\theta^{2}+sin^{2}\theta d\varphi^{2})-
e^{-2\delta(r)}H(r)dt^{2}
\end{equation}
where $H(r)=(1-2m(r)/r-\Lambda r^2/3)$ and $m(r)$ may be interpreted  
as the total  mass-energy within the radius $r$. 
For black hole solutions, the event horizon is at $r=r_{h}$  where  $H(r_{h})=0$.

Following \cite{10}, we assume that $\Phi$ posseses
 only one degree of freedom
\begin{equation}
\Phi=\frac{1}{\sqrt{2}}{0 \choose \phi(r)},
\end{equation}
with $\phi$ real, and has the standard double-well Higgs potential
\begin{equation}
V(\phi)=\frac{\lambda}{4}(\phi^{2}-v^{2})^{2},
\end{equation}
where $v$ denotes the vacuum expectation value of $\Phi$. 

A suitable parametrization of the Yang-Mills connection is \cite{10}
\begin{equation} \label{sch2}
A=\frac{1}{g}(1+\omega(r))[-\widehat\tau_{\varphi}d\theta+\widehat\tau_{\theta} \sin \theta d\varphi]
+\frac{1}{g}a_0(r)\widehat\tau_{r}dt.
\end{equation}
The $\widehat\tau_{i}$ are appropriately normalised spherical generators of the SU(2) group in 
the notation of ref. \cite{10}, $e.g.$ $\widehat\tau_{r}=\widehat r\cdot\tau$, 
$[\tau_{a},\tau_{b}]=\epsilon_{abc}\tau_{c}$, while $\omega (r)$ and $a_0(r)$ are the magnetic and 
electric YM potentials for a spherically symmetric configuration. 
\subsection{A no-dyons theorem}
When discussing Einstein-SU(2) system in asymptotically Minkowski spacetimes there are 
no-go theorems forbidding the electric 
components of the gauge field for finite energy static solutions \cite{14, 15}. This result
is also valid in EYMH theory \cite{10} with a doublet Higgs field.
If we consider a negative cosmological constant, these no-go theorems 
fail for the EYM system \cite{5},
thus permitting dyon solutions.
However, this fact does not generalize when including a Higgs field. 
This can be proven by using the equation for $a_0(r)$
\begin{equation}
(a_0'r^{2} e^{\delta})'= 2 a_0 \omega^2 \frac{e^\delta}{H}+a_0
 (\frac{g \phi r}{2})^2 \frac{e^\delta}{H}
\end{equation}
(where the prime denotes derivative with respect to $r$)
which implies an asymptotic solution consistent with the finite-energy constraints
on the form
\begin{equation}
a_0(r)\sim \frac{c}{r^{\frac{1}{2}(1+\sqrt{1-3(g v)^2/\Lambda})}}
\end{equation}
(note that a nonvanishing value of Higgs field at infinity forces $a \rightarrow 0$ and  
implies a different behavior compared to the pure EYM system).
From the field equations we derive the sum rules
\begin{eqnarray}
\frac{1}{2}((a_0^2)'r^2 e^{\delta})'= 
[2 a_0^2 \omega^2+(\frac{g \phi a_0 r}{2})^2]\frac{e^\delta}{H}
+(a_0')^2r^2 e^{\delta},
\nonumber \\
(H e^{-\delta} \frac{\omega'}{\omega})'=\frac{(\omega^2-1) e^{-\delta}}{r^2}
+\frac{\omega +1}{4\omega}e^{-\delta} (g\phi)^2-a_0^2\frac{e^\delta}{H}
-(\frac{\omega'}{\omega})^2 H e^{-\delta}.
\end{eqnarray}
By integrating these expresions from $r_0$ to $\infty$ (where $r_0=0$ or $r_h$)
we conclude that $a_0=0$ for finite energy solutions.

\subsection{Reduced action and virial relations}
Expressing  the curvature  scalar $\mathcal{R}$ in terms of the metric  function  $m(r)$ and
$\delta(r)$, we obtain the following  expression of the effective action of our static
spherically symmetric system:
\begin{eqnarray}
S=\int dr e^{-\delta}
[\frac{m'}{G}-\frac{1}{2}\phi'^2 r^2 (1-\frac{2 m}{r}-\frac{\Lambda r^2}{3})
-(\frac{\phi}{2})^2(1+\omega)^2 - V r^2
\nonumber \\
-\frac{1}{g^2}(\omega')^2(1-\frac{2 m}{r}-\frac{\Lambda r^2}{3})
-\frac{1}{g^2}\frac{(1-\omega^2)^2}{2 r^2}
].
\end{eqnarray}

A  usual  rescaling  \cite{16}
\begin{eqnarray} \label{rescaling}
r\to rg/\sqrt{G},
\quad 
\phi\to\phi/\sqrt{G}
\end{eqnarray}  
 reveals  the existence of two  dimensionless  parameters $\alpha$ and $\beta$, 
 expressible  through the mass
ratios
\begin{eqnarray} 
\alpha=\frac{M_{W}}{M_{Pl}};
\quad
 \beta=\frac{M_{H}}{M_{W}}
\end{eqnarray}
with $M_W=gv$, $M_H=\sqrt{\lambda}v$ and $M_{Pl}=\frac{1}{\sqrt{G}}$ . 
The third parameter of the system is the rescaled cosmological constant 
$ \Lambda\to \Lambda G/g^2$.
The scalar field potential after rescaling is 
$V(\phi)=\frac{\beta^{2}}{4}(\phi^{2}-\alpha^{2})^{2}$.

This form of the reduced action allow us to obtain some interesting virial relations. 
In this way it is possible to better understand the reason for the existence of 
nontrivial solutions in EYM or EYMH system and to provide nonexistence theorems 
for the Einstein-scalar field system.

We will use the approach proposed by Heusler in \cite{17, 18}
(although similar results can be obtained by using
 a curved spacetime version of Deser's argument \cite{7}). 
Let us assume the existence of a solution $ m(r), \delta(r), \omega(r), \phi(r)$
 with suitable boundary conditions at the origin and at infinity. 
 Then each member of the 1-parameter family
\begin{equation}
m_\lambda(r) \equiv m(\lambda r), \ 
\delta_{\lambda} (r) \equiv \delta (\lambda r), \ 
\omega_{\lambda} (r) \equiv \omega (\lambda r), \ 
\phi_{\lambda} (r) \equiv \phi (\lambda r)
\end{equation}  
assumes the same boundary values at $r=0$ and $r=\infty$, and the action 
$S_{\lambda} \equiv S[m_{\lambda}, \delta_{\lambda}, \omega_{\lambda},\phi_{\lambda}]$
must have a critical point at $\lambda=1$, $[dS/d\lambda]_{\lambda=1}=0$.
Thus   we obtain the virial relation valid for a regular spacetime
\begin{eqnarray} \label{virial1}
\int_{0}^{\infty}  e^{-\delta}[(\omega')^2 (1-\frac{4 m}{r}
+\frac{\Lambda r^2}{3}) 
+\frac{1}{2r^2}(1-\omega^2)^2 ]dr= 
\nonumber \\
\int_{0}^{\infty}  e^{-\delta}[\frac{1}{2}\phi'^2 r^2
(1-\Lambda r^2) 
+\frac{1}{4}\phi^2 (1+\omega)^2
+3Vr^2]dr.
\end{eqnarray}
For a black-hole spacetime, a suitable parametrization
\begin{eqnarray}
m_\lambda(r) \equiv m(r_h+\lambda (r-r_h)), \
\delta_{\lambda} (r) \equiv \delta (r_h+\lambda (r-r_h)),
\nonumber \\
\omega_{\lambda} (r) \equiv \omega (r_h+\lambda (r-r_h)), \ 
\phi_{\lambda} (r) \equiv \phi (r_h+\lambda (r-r_h))
\end{eqnarray}
implies the identity
\begin{eqnarray} 
\int_{r_h}^{\infty}  e^{-\delta}[
(\omega')^2 (1+\frac{2m}{r}(\frac{r_h}{r}-2)+
\frac{\Lambda r^2}{3}(1-\frac{2 r_h}{r}))
+\frac{(1-\omega^2)^2}{2 r^2}(1-\frac{2r_h}{r})
\nonumber \\
+\frac{1}{2}\phi'^2 r^2
(\frac{2 r_h}{r}(1-\frac{m}{r})-1-\frac{\Lambda r^2}{3}(\frac{4r_h}{r}-3))
-\frac{\phi^2}{4}(1+\omega)^2+Vr^2(\frac{2r_h}{r}-3)]dr=0.
\end{eqnarray}
It is possible in this way to better understand the existence of 
stable selfgravitating EYM soliton and black hole solutions (without a scalar field).
For a fixed AdS background ($i.e.$ neglecting 
the backreaction) we have the relation
\begin{eqnarray} 
\int_{0}^{\infty}  [(\omega')^2 (1+\frac{\Lambda r^2}{3}) 
+\frac{1}{2r^2}(1-\omega^2)^2 ]dr=0.
\end{eqnarray} 
Thus, different from Minkowski spacetime one can not use a scaling argument
to exclude the existence of finite energy, nontrivial YM configurations.
This fact suggests the existence of finite energy solutions of YM
equations in a AdS background (note the absence of these solutions for 
$\Lambda \geq 0$). 

A nontrivial exact solution of YM equations 
(noticed in \cite{Boutaleb-Joutei:1979va} for a positive cosmological constant) is
\begin{eqnarray} \label{exact}
\omega=1/(1-\Lambda/3 r^2)^{1/2},
\end{eqnarray} 
describing a monopole in AdS spacetime with unit magnetic charge 
and mass $\sqrt{(-3\Lambda)} \pi/8$.
A preliminary numerical study gives us 
strongly numerical arguments for the existence of an entire family of solutions 
((\ref{exact}) being a particular case)
 with very similar properties with the
selfgravitating counterparts. 
A study of these solutions will be presented elsewhere.

By using this relation we can also discuss whether the cosmological constant 
can support a real scalar field.
In asymptotically flat spacetimes, an well-known result implies the absence of 
scalar solitons. 
Also, there are different proofs of the no-hair theorem for spherically
symmetric scalar fields (see \cite{17} for a set of references). 

The nonexistence of real static solutions is partially due to the boundary conditions
at the origin and at infinity. 
Hence we expect that the previous results may be changed by the presence of a 
cosmological constant which implies a different asymptotic structure of spacetime.  

This is the case for $\Lambda>0$; in ref. \cite{19} it was shown that a positive 
cosmological constant can support a selfinteracting scalar field  
and specific (unstable) solutions were exhibited.

However, when $\Lambda<0$, the situation is similar to
the asymptotically flat one.
For the Einstein-scalar field theory with arbitrary nonnegative potential
 we can use (\ref{virial1}) to exclude the existence
of spherically symmetric scalar solitons:
\begin{eqnarray}  
\int_{0}^{\infty}  e^{-\delta}[\frac{1}{2}\phi'^2 r^2
(1-\Lambda r^2)
+3Vr^2]dr=0.
\end{eqnarray} 
For a black hole solution we have
\begin{eqnarray} 
\int_{r_h}^{\infty}  e^{-\delta}[\frac{1}{2}\phi'^2 r^2
(\frac{2 r_h}{r}(1-\frac{m}{r})-1-\frac{\Lambda r^2}{3}(\frac{4r_h}{r}-3))
+Vr^2(\frac{2r_h}{r}-3)]dr=0.
\end{eqnarray}
Since the factors of $\phi'^2$ and $V(\phi)$ are strictly decreasing negative 
quantities, we conclude the absence of black holes with scalar hair.
\subsection{Field equations and boundary conditions}
The field equations implies the relations
\begin{eqnarray} \label{weq}
[1-\frac{2 m}{r}-\frac{\Lambda r^2}{3}]\omega''+
[\frac{2 m}{r^2}-\frac{2\Lambda r}{3} -
\frac{\phi^2(1+\omega)^2}{r}
\nonumber \\
-2V r -\frac{(1-\omega^2)^2}{r^3}]\omega'
-\frac{\phi^2 (1+\omega)}{4}
-\frac{\omega (\omega^2-1)}{r^2}=0
\end{eqnarray}
for the gauge field, and
\begin{eqnarray} \label{higseq}
[1-\frac{2 m}{r}-\frac{\Lambda r^2}{3}]\phi'' r^2
+[2r-2m-\frac{4 \Lambda r^3}{3}
-\frac{\phi^2 (1+\omega)^2 r}{2}
\nonumber \\
-2V r^3-\frac{(1-\omega^2)^2}{r}]\phi'
-\frac{\phi (1+\omega)^2}{2}-\frac{dV}{d\phi}r^2=0
\end{eqnarray}
for the Higgs field.
 The $(rr)$ and $(tt)$ Einstein equation are
\begin{eqnarray} \label{meq}
m'=
(\frac{\phi'^2 r^2}{2}+\omega'^2)
(1-\frac{2 m}{r}-\frac{\Lambda r^2}{3})+\frac{\phi^2 (1+\omega)^2}{4}
+Vr^2+\frac{(1-\omega^2)^2}{2 r^2},
\end{eqnarray}
\begin{eqnarray} \label{deltaeq}
\delta'=-\frac{2}{r}(\omega'^2+\frac{\phi'^2 r^2}{2}).
\end{eqnarray}
Following the analysis in \cite{10}, we can predict the 
boundary conditions and some general features of the finite energy solutions.
When discussing the pure EYM system (with $\Lambda<0$), there are no restriction 
for the asymptotic values of $\omega$.

However, in the presence of a Higgs field, the boundary conditions at infinity 
obtained for an asymptotically flat spacetime remain valid and  
 $\omega=-1$ is the  only
acceptable value.
Also $\pm\alpha$ are the only  allowed  values of $\phi$ 
as $r\to\infty$; 
we focus  here on  solutions  with 
  $\phi(\infty)=\alpha$ without loss of generality. 
The vacuum values $\omega(\infty)=-1$ and 
  $\phi(\infty)=\alpha$    
are shared by both black-holes and regular solutions. 
The analysis of the field equations as $r\to\infty$ gives
\begin{eqnarray} \label{masimpt}
m(r) &\sim & M +
\frac{(\frac{\Lambda(1+k_1)^2}{3}-
\frac{\alpha^2}{4})}{1+2k_1}\frac{c^2_{1}}{r^{1+2k_1}},
\nonumber \\
\omega &\sim & -1+\frac{c_1}{r^{1+k_1}},
\nonumber \\
\phi &\sim & \alpha-\frac{c_2}{r^{3+k_2}},
\nonumber \\
\delta &\sim & \frac{(1+k_1)^2}{2+k_1}\frac{c^2_1}{r^{4+2k_1}},
\end{eqnarray}
where $k_1=\frac{1}{2}(\sqrt{1-3 \alpha^2/\Lambda}-1)>0$, 
$k_2=\frac{3}{2}(\sqrt{1-8 \alpha^2 \beta^2/3\Lambda}-1)>0$; 
$M, c_1, c_2$ are positive constants ($M$ being the total mass of the configuration).
Thus, the Higgs field implies a complicated power decay at infinity,
rather than polynomial (the case of EYM fields) or exponentially (EYMH theory in
an asymptotically flat space).
Note also the absence of magnetic charge implied by these boundary conditions.

 By using the relation
\begin{eqnarray}
\frac{1}{2}((\omega^2)'H e^{-\delta})'=
\frac{e^{-\delta}}{r^2}((\omega^2-1)\omega^2
+(\frac{\phi}{2})^2(1+\omega)\omega r^2)
+\omega'^2 H e^{-\delta}
\end{eqnarray}
we find that $\omega\leq1$ is  required  for finite  energy  solutions.
This constraint is valid for both regular and black-hole spacetimes.
The Higgs equation can be rewritten in the form
\begin{eqnarray}\label{fi2}
\frac{1}{2}((\phi^2)'r^2 e^{-\delta})'=
(r\phi')^2 e^{-\delta}
+\frac{e^{-\delta}}{H}(\frac{(1+\omega)^2\phi^2}{2 r^2}
+\phi\frac{dV}{d \phi}).
\end{eqnarray}
The obvious requirement for finite energy solution is
\begin{eqnarray} \label{rel1}
\phi\frac{dV}{d\phi}<0
\end{eqnarray}
which implies that $\phi$ is restricted to lie between the minima of the potential,
$-\alpha\leq\phi\leq\alpha$.        
\section{NUMERICAL SOLUTION}

\subsection{Regular solutions}
Similar to the asymptotically Minkowski spacetime case, we have found 
two possible set of initial conditions for 
regular solutions

\begin{eqnarray} 
2m(r)=O(r^{3})
\\
\delta(r)=O(r^{2})
\\
{\omega (r) \choose \phi (r)}={-1+O(r^{2}) \choose \phi_{0}+O(r^{2})},
\end{eqnarray}
or 
\begin{eqnarray}
{\omega (r) \choose \phi (r)}={1+O(r^{2}) \choose O(r)}.
\end{eqnarray}
The general properties of the solutions are the same as for the $\Lambda=0$ case.
  Solutions  are again  characterized  by $\omega(r) $  oscillations  in the
region $r>1$ and classified by the node number $k$ which may be even or odd.  The
formal power series describing the above boundary conditions at $r=0$ is
\begin{eqnarray} 
2m(r)&=&[4b^2+\frac{2}{3}V_0]r^{3}+
\frac{2}{5}[-8b^3+(\frac{3\phi_0^2}{4}+\frac{16}{3}V_0-\frac{4\Lambda}{3})b^2
\nonumber\\
&&+\frac{2}{9}(V_0')^2]r^{5}+O(r^{7}),
\\
\delta(r)&=&-4br^{2}-[\frac{V_0'}{6}
+\frac{4b^2}{5}(8b^2-3b+\frac{\phi_0^2}{4}+4V_0+2\Lambda)]r^4+O(r^{6}),
\\
\omega&=&-1+br^{2}+\frac{1}{10}[8b^3-3b^2+(\frac{\phi_0^2}{4}+4V_0+2\Lambda)b]r^{4},
\\
\phi&=&\phi_{0}+\frac{1}{6}V'_0r^{2}+[(\frac{1}{40}\phi_0+\frac{9}{45}V_0')b^2
\nonumber\\
&&+V_0'(\frac{1}{18}V_0+\frac{1}{120}V_0''+\frac{1}{36}\Lambda)]r^4+O(r^{6}),
\end{eqnarray}
for  even-k  solutions  ($V_0, V'_0,V''_0$   are the  potential  and its  derivatives  with
respect to $\phi$ at $\phi=\phi_0$) and
\begin{eqnarray}
2m(r)&=&[4b^{2}+\frac{2}{3}V_{0}+e^{2}]r^{3}+\frac{2}{5}
[-8b^3+\frac{8b^2}{3}(2V_0+\Lambda)
\nonumber\\
&&+e^2(6b^2-3b+e^2+V_0+V_0''+\frac{\Lambda}{2})]r^5+O(r^{7}),
\\
\delta(r)&=&-[4b^{2}+\frac{e^2}{2}]r^{2}+\frac{1}{5}
[12b^3-32b^2+\frac{7be^2}{2}-22b^2e^2
\nonumber\\
&&-\frac{9e^4}{4}+\frac{3e^2V_0''}{4}
-(8b^2+e^2)(2V_0+\Lambda)]r^4+O(r^{6}),
\\
\omega &=&1-br^{2}-\frac{1}{10}[8b^3-3b^2
+2b(2V_0-2e^2+\Lambda)-\frac{e^2}{2}]r^4+O(r^{6})
\\
\phi&=&er+\frac{e}{10}[8b^2-2b+3e^2+\frac{8V_0}{3}+V_0''+\frac{4\Lambda}{3}]r^3+O(r^{5}),
\end{eqnarray}
for odd-k solutions. 

Given ($\alpha, \beta, \Lambda$), 
solutions may exist for a discrete set of shooting parameters
 $(\phi_0,b)$ and $(e,b)$ respectively.
We follow the usual approach and, by using a standard ordinary  differential  
equation solver, we
evaluate  the  initial  conditions  at  $r=10^{-3}$ for  global  tolerance  
$10^{-12}$, adjusting  for fixed shooting parameters and  integrating  towards  $r\to\infty$.
The  difficulty  of the two-dimensional  shooting problem in the presence of three 
free parameters is increased by the asymptotic power law decay 
which leads to a slow convergence of the gauge function $\omega (r)$.

In the vicinity of origin, these solutions resemble the solutions 
found by Greene, Mathur and O'Neill.
The asymptotic behavior is however somewhat different.
We have started our numerical investigation by considering the case $\Lambda=0$ and slowly 
increasing the value of cosmological constant for fixed ($\alpha, \beta$).

The results obtained for the $k=1$ and $k=2$ solutions retains 
the general characteristics of the $\Lambda=0$ case: two
solutions brances again for each $k$, distinguished by a different behavior in the limit
 $\alpha\to 0$.
In this limit (corresponding to a vanishing Newton constant), for the quasi- $k=0$ branch, 
we have weakly coupled gravity in a nonflat geometry and the solution approaches the standard model sphaleron in a
AdS spacetime background, while the node moves to infinity
 (an explanation of this behavior is given in \cite{1, 10}) .

In the same limit (corresponding this time to a vanishing Higgs field) 
the proper- $k=1$ branch approaches a particular one node regular solution
from the family of solutions discussed in \cite{4, 5}.

For two nodes solutions we have noticed a similar behavior. There are again two branches, 
quasi- $k=1$ and proper- $k=2$ with a different behavior as  $\alpha\to 0$.
In this limit, the proper-  $k=2$ branch approaches a particular two-nodes EYM regular solution, while 
the proper-  $k=2$ approches the corresponding one-node regular solution, 
the second node moving to infinity. 

However, similar to the $\Lambda=0$ case we expect these branches to converge 
for suitable values of $\alpha, \beta$ \cite{1, 11}.  
 
A complete analysis of the complex correlation between the three parameters
of the theory ($\alpha, \beta, \Lambda$) is beyond the purposes of this paper.
  To compare numerically the results with those found in \cite{10} we 
focused on solutions with $\beta^{2}=1/8$ and have varied the parameter $\alpha$ for a 
limited set of $\Lambda$.
Some results of the numerical  integration are  presented in $figure$ 1a-d.

Significant differences occur for large enough negative values 
of $\Lambda$; the  parameter  range obtained in \cite{10} for the two sheets of 
solutions does not remain  valid.
In particular we have found a different maximal value of $\alpha$ for every value of 
$\Lambda$, such that above 
$\alpha_{max}$ the solution ceases to exist.
As $|\Lambda|$ increases, the mass of the solution increases also, while 
the value of $\alpha_{max}$ decreases.

For example, for the quasi- $k=1$ branch, Greene, Mathur and O'Neill \cite{6}
 have found
$0<\alpha<0.122$; when $\Lambda=-0.001$ we have obtained $0<\alpha<0.115$.
For the proper- $k=1$ branch the solution ceases to exist for $\alpha>0.454$ ($\Lambda=-0.1)$,
 while as $\Lambda \to 0$ it has been found in \cite{6} that $0<\alpha<0.599$ .

Conversely, we have noticed a maximal allowed value of $\Lambda$ for a given value of 
$\alpha<\alpha_{max}${\small ($\Lambda=0$)}. For example, for the proper- $k=1$ branch
when $\alpha=0.1$  we have found solutions for $|\Lambda|<0.35$ only, 
while for proper- $k=2$ branch and the same value of $\alpha$,  $|\Lambda|_{max} \simeq 0.0032$. 

Different limiting values occur for the shooting 
parameters $b, \phi_{0}$ and $e$ also.  
However, given the asymptotic behaviour (\ref{masimpt}) to find accurate 
values for limiting parameters is a more difficult problem compared to the $\Lambda=0$ case.


\subsection{BLACK HOLE SOLUTIONS}
Similar  results can be obtained for numerical  black hole  solutions.  We use
the following expansion near the event horizon:
\begin{eqnarray} \label{m(rh)}
m(r)=\frac{r_{h}}{2}-\frac{\Lambda}{6}r_h^3+m'(r_{h})(r-r_{h}),
\end{eqnarray}
\begin{eqnarray} \label{delta(rh)}
\delta(r)=0+\delta '(r_{h})(r-r_{h}),
\end{eqnarray}
\begin{eqnarray} \label{w(rh)}
\omega (r)=\omega (r_{h})+\omega '(r_{h})(r-r_{h}),
\end{eqnarray}
\begin{eqnarray} \label{fi(rh)}
\phi (r)=\phi (r_{h})+\phi '(r_{h})(r-r_{h}),
\end{eqnarray}
with
\begin{eqnarray}
&&m'(r_{h})=(\frac{\phi(r_h)}{2})^2(1+\omega(r_h))^2
+V(\phi(r_h))r^2_h+\frac{(1-\omega^2(r_h))^2}{2r_h^2},
\\
&&\omega '(r_{h})=\frac{(\phi(r_h)/2)^2(
1+\omega(r_h))+\omega(r_h)(\omega(r_h)^2-1)/r_h^2}
{1/r_h-\Lambda r_h-\frac{\phi^2(r_h)(1+\omega(r_h))^2}{2r_h}
-2V(\phi(r_h))r_h-\frac{(1-\omega(r_h)^2)^2}{r_h^3}},
\\
&&\phi '(r_{h})=\frac{\phi(r_h)(1+\omega(r_h))^2/2+V'(\phi(r_h))r_h^2}
{r_h-\Lambda r_h^3-\frac{\phi(r_h)^2(1+\omega(r_h))^2r_h}{2}-2V(\phi(r_h))r_h^3
-\frac{(1-\omega(r_h)^2)^2}{r_h}},
\\
&&\delta '(r_{h})=-(2\omega'(r_h)^2+\phi'(r_h)^2r_h^2)/r_h.
\end{eqnarray}
The new shooting parameters are $\omega (r_{h})$ and  $\phi(r_{h})$ (we have focused on the 
case $r_{h}=1$ only).

Starting from the solutions (\ref{m(rh)}-\ref{fi(rh)}) we integrated the system 
(\ref{weq}-\ref{deltaeq}) towards $r\to\infty$ using an automatic
step procedure and accuracy $10^{-12}$. The integration stops when the AdS spacetime asymptotic
limit (\ref{masimpt}) is reached.
 
Similar to the regular solutions case, two solution branches appear for each
$k$. As
$\alpha\to 0$, the proper- $k=1, 2$ branches approaches the corresponding particular 
cases of the
EYM  black-hole solutions discussed in ref. \cite{5, 6}.
In the same limit, the quasi- $k=0$ branch is distinguished by 
its Schwarzschild-anti-de Sitter solution limit
 ($\omega =1$, $\phi=0$) and the last node of the  quasi- $k=0$ and  quasi- $k=1$
 branches is again pushed out to infinity.

The results for $k=1, 2$, $\beta^{2}$=1/8 and various values of 
the parameters $\Lambda, \alpha$ are presented in $figure$ 2.
As expected, for a nonzero $\Lambda$ it is necesary to establish new limiting values of the 
values of the normalised vacuum expectation values $\alpha$. For example, an asymptotically flat  
solution has necessarily 
$0<\alpha<0.356$ (quasi- $k=0$ branch); when $\Lambda=-0.01$ we have found that
for $\alpha>0.323$ solution ceases to exist.
 For $\Lambda=-0.001$ the allowed range of $\alpha$ is
 $0<\alpha<0.029$ (proper- $k=2$ branch) while while for ($\Lambda = 0)$,   $0<\alpha<0.047$. 

As a general feature,
 we have noticed a decresing of the maximal 
allowed value of the parameter $\alpha$ and a larger ADM mass for fixed ($\alpha, \beta$)
compared with $\Lambda=0$ case. 
Different ranges for the shooting parameters $\omega (r_{h})$ and  $\phi(r_{h})$ are 
to be imposed. 

For the considered values of ($\alpha, \Lambda$) (with $\beta^2=1/8$) we did not noticed 
the occurence of critical solutions existing in pure EYM theory \cite{5} ($i.e.$ for $ r=r_c$ 
$H(r_c)=0,\ e^{\delta(r_c)}=0$). This fact is valid for both regular and black hole solutions. 

\section{STABILITY ANALYSIS}
An important physical question when discussing selfgravitating 
nonlinear field configurations is whether these solutions are stable.

Although the instability of the sphaleron 
sector is expected for topological reasons,
we have addressed this problem for our system motivated by recent results in EYM theory in
 asymptotically AdS spacetime \cite{4, 5}. Also, when the spacetime is not asymptotically flat
 the stability analysis can be a quite involved and subtle problem, mainly
for topological reasons \cite{3}.
Since we have not exhaustively studied this question, 
we briefly discuss here only the issue  of stability of the soliton solutions.

Boschung et. al. used a powerful method to prove the instability of 
asymptotically flat gravitating sphalerons, by studying the frequency spectrum
of a class of radial perturbations \cite{12}. 
With the help of a variational principle they have shown
that there are always unstable modes. The same method will now be used to show 
that the solutions described in section $(3.1)$
are also unstable. This method has the advantage that no detailed knowledge 
of the equilibrium solution is required.

Since the proof is practically similar to that valid for $\Lambda=0$, 
we will present here the main steps only, emphasizing the points where 
the different asymptotic structure of spacetime is  essential.

When considering spherically symmetric perturbations of the regular solution, 
the metric is time-dependent
\begin{equation}
ds^{2}=\frac{dr^{2}}{N(r,t)}+r^2(d\theta^{2}+\sin^{2}\theta d\varphi^{2})-
S^2(r,t)N(r,t)dt^{2},
\end{equation}
(where $N(r,t)=1-2m(r,t)/r-\Lambda r^2/3$), while following  \cite{12} 
we assume the following ansatz for 
the non-abelian gauge potential
\begin{eqnarray}
A&=&a_0(r,t)\widehat\tau_{r}dt+a_1(r,t)\widehat\tau_{r}dr
+(1-\omega(r,t))[-\widehat\tau_{\varphi}d\theta+\widehat\tau_{\theta} \sin \theta d\varphi]
\nonumber\\
&^{}&+\tilde{\omega}(r,t)[\widehat\tau_{\theta}d\theta+\widehat\tau_{\varphi} \sin \theta d\varphi]
\end{eqnarray}
(note that we have to redefine ($\omega \to - \omega$) in (6)
 to conform with conventions of \cite{12, 13}).
Also, the spherically symmetric Higgs field has the form
\begin{equation}
\Phi=\frac{1}{\sqrt{2}}{0 \choose \phi(r,t)}+\frac{1}{\sqrt{2}}\psi(r,t) \widehat\tau_r.
\end{equation}
The solutions considered in Section 3.1 have nonvanishing $H(r),N(r),w(r)$ and $\phi(r)$
but $a_1=\tilde{\omega}=\psi$=0.

In examining time-dependent perturbations around gravitating 
sphaleron solutions it is convenient 
to work in the $a_0=0$ gauge. 
Thus, the linearized perturbation equations decouple into two sectors \cite{12}.
The first consist of the gravitational modes $\delta N, ~\delta S,
~\delta \omega$ and $\delta \phi$ and the second of the matter perturbations
$\delta a_1,~\delta \tilde{\omega}$ and $\delta \psi$. 
We are interested in the matter perturbations only, 
because we shall find instabilities within this class.
In this case the metric perturbations $\delta m$ and $\delta S$ vanish identically.
For a harmonic time dependence $e^{i \Omega t}$ the linearized system 
(with respect to these perturbations) obtained from the matter 
equations of motion can be cast into a Schr\"odinger eigenvalue problem
\begin{equation} \label{sch1}
\mathcal{H} \Psi=\Omega^2 \mathcal{A} \Psi.
\end{equation}
The explicit expression of operators $\mathcal{H}, \mathcal{A}$ is given in \cite{12} (rels. (23-33)) 
(see also \cite{13, 11}),
while
\begin{displaymath}
\Psi=\left( \begin{array}{cc} \delta a_1\\
\delta \tilde{\omega}
\\
\delta \psi
\end{array}
\right)
\end{displaymath}
(this time we have to use the metric functions $N=H$ and $S= e^{-\delta}$).
It can be shown that $\mathcal{H}$ is a self-adjoint operator with respect to the inner (scalar) 
product in the space of functions ${\Psi}$, and the $\mathcal{A}$ matrix is positive 
definite $<\Psi|\mathcal{A}|\Psi>>0$.
A criterium for instability is the existence of an imaginary frequency mode in (\ref{sch1}).
We make use of the following functional defined as
\begin{equation}
\Omega^2(\Psi)=\frac{<\Psi|\mathcal{H}|\Psi>}{<\Psi|\mathcal{A}|\Psi>},
\end{equation}
where $\Psi$ is this time a trial function. The criteron of instability in this approach reads
\begin{eqnarray}
\Omega^2(\Psi)<0,
\nonumber\\
<\Psi|\mathcal{A}|\Psi> < \infty.
\end{eqnarray}
An essential point in our proof is the fact that for $\Lambda<0$ the matter 
functions satisfy the same boundary conditions at the origin and at infinity as in the  
$\Lambda=0$ case.

Thus we may choose as trial perturbations the following expressions \cite{12}
\begin{eqnarray}
\delta a_1&=&\omega',
\nonumber\\
\delta \chi&=&\omega^2-1,
\nonumber\\
\delta \xi&=&\frac{(\omega-1) \phi}{2},
\end{eqnarray}
where ($\omega, \phi$) is the unperturbed solution discussed in section (3.1).
The expressions $<\Psi|\mathcal{A}|\Psi>$ and $<\Psi|\mathcal{H}|\Psi>$ computed in \cite{12} for a 
general $N(r)$, $S(r)$ are
\begin{eqnarray}
<\Psi|\mathcal{A}|\Psi>&=&\int{\{\frac{r^2(\omega')^2}{S}+2\frac{(\omega^2-1)^2}{NS}
+\frac{(\omega-1)^2\phi^2r^2}{4NS}\}}dr,
\nonumber\\
<\Psi|\mathcal{H}|\Psi>&=&-\int{ \{ 2N(\omega')^2+2\frac{(\omega^2-1)^2}{r^2}
+\frac{(\omega-1)^2\phi^2}{2}\} }dr<0.
\end{eqnarray}
Although for a $\Lambda<0$ the unperturbed solution 
does no longer exponentially vanish as $r \to \infty$ 
the relations (\ref{masimpt}) still assure the finiteness of $<\Psi|\mathcal{A}|\Psi>$.

Therefore $\Omega^2$ is clearly negative and the instability of the gravitating sphaleron
solution is proven.

This instability has been found in the sphaleron sector of the theory;
 a crucial point was the fact that we did not use the Einstein equations,
nor the explicit form of the metric functions (with a $\Lambda r^2$ term).
 
The case of sphaleron black hole solutions is somewhat different.
Although the corresponding asymptotically flat configurations have been proven to be unstable, 
the aproach used in \cite{11, 13} can not be directly applied  for an asymptotically AdS spacetime.
However it seems unlikely that they are stable.


\section{CONCLUSIONS}

In this work we have analysed the basic properties of gravitating sphaleron 
and black hole solutions of the $(3+1)-$dimensional EYMH system in the presence of a negative
cosmological constant.

Both analytical and numerical arguments have been presented for the existence of nontrivial
solutions. A general virial relation has been found and particular cases  have been discussed.
An analytic solution of the Yang-Mills equations in fixed anti-de 
Sitter background has been noticed.
For the Einstein-minimally coupled scalar field system with a positive scalar field potential
we have shown the absence of both regular and black hole solutions when $\Lambda<0$.

The numerical solutions of the full EYMH system we have found do not retain the 
nontrivial properties of 
the pure EYM case in an asymptotically AdS spacetime and are rather similar 
to those corresponding to an asymptotically flat spacetime
 
(a similar behavior has been found for the monopole and dyon
 solutions of an EYMH theory with the Higgs field 
in the adjoint representation).

Thus, it seems that when studying the EYMH system
the presence of the scalar field induces a kind of generic behavior of the solutions, 
valid for $\Lambda \leq 0$ (given the presence of a cosmological event horizon, 
the case $\Lambda>0$ needs a separate analysis).

However, a nonzero $\Lambda$  term in the action implies a power decay of the fields
at infinity and a decrease of the 
maximal allowed vacuum  expectation value of the Higgs field.
Also, the mass of the solutions for a fixed vacuum expectation value of the Higgs field
 has a greater value
compared to the $\Lambda=0$ case.

Given the similarities with the EYMH case, we expect to obtain a very similar behavior 
for the solutions of a Non-Abelian-Proca theory.
\\
\newline
{\bf Acknowledgement}
\newline
 This work was performed in the context of the
Graduiertenkolleg of the Deutsche Forschungsgemeinschaft (DFG):
Nichtlineare Differentialgleichungen: Modellierung,Theorie, Numerik, Visualisierung.


\newpage
{\bf Figure Captions}
\newline

Figure 1:
One- and two-node sphaleron solutions of the EYMH theory
 for $\beta^{2}=1/8$ and various values of $\Lambda$, $\alpha$.
\newline 
\\
\\
\\

Figure 2:
One- and two-node black-hole solutions of the EYMH theory
 for $\beta^{2}=1/8$ and various values of $\Lambda$, $\alpha$.
\\

\newpage
\setlength{\unitlength}{1cm}

\begin{picture}(16,16)
\centering
\put(-2,0){\epsfig{file=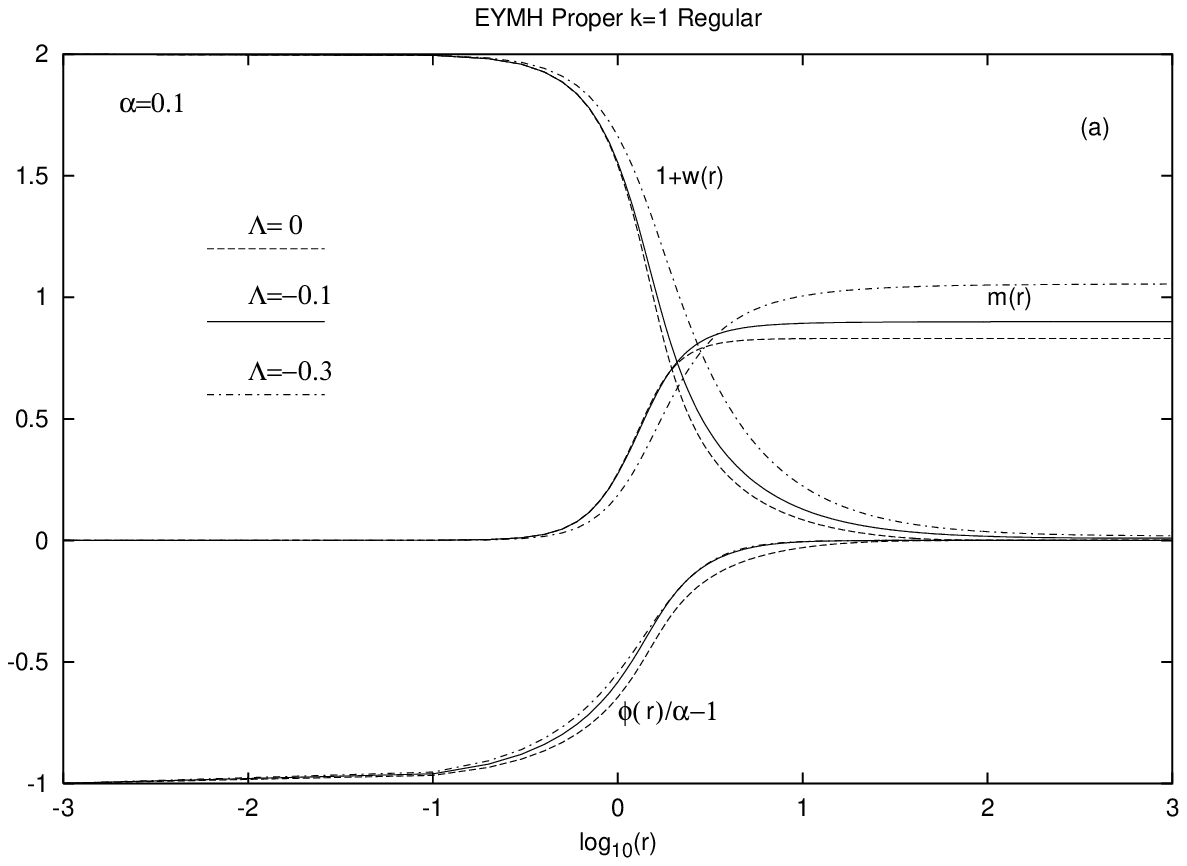,width=16cm}}
\end{picture}
\begin{center}
Figure 1a.
Proper $k=1$ regular $\alpha=0.1$; { }$\Lambda=0$, -0.1, -0.3\newline
\end{center}

\newpage
\begin{picture}(16,16)
\centering
\put(-2,0){\epsfig{file=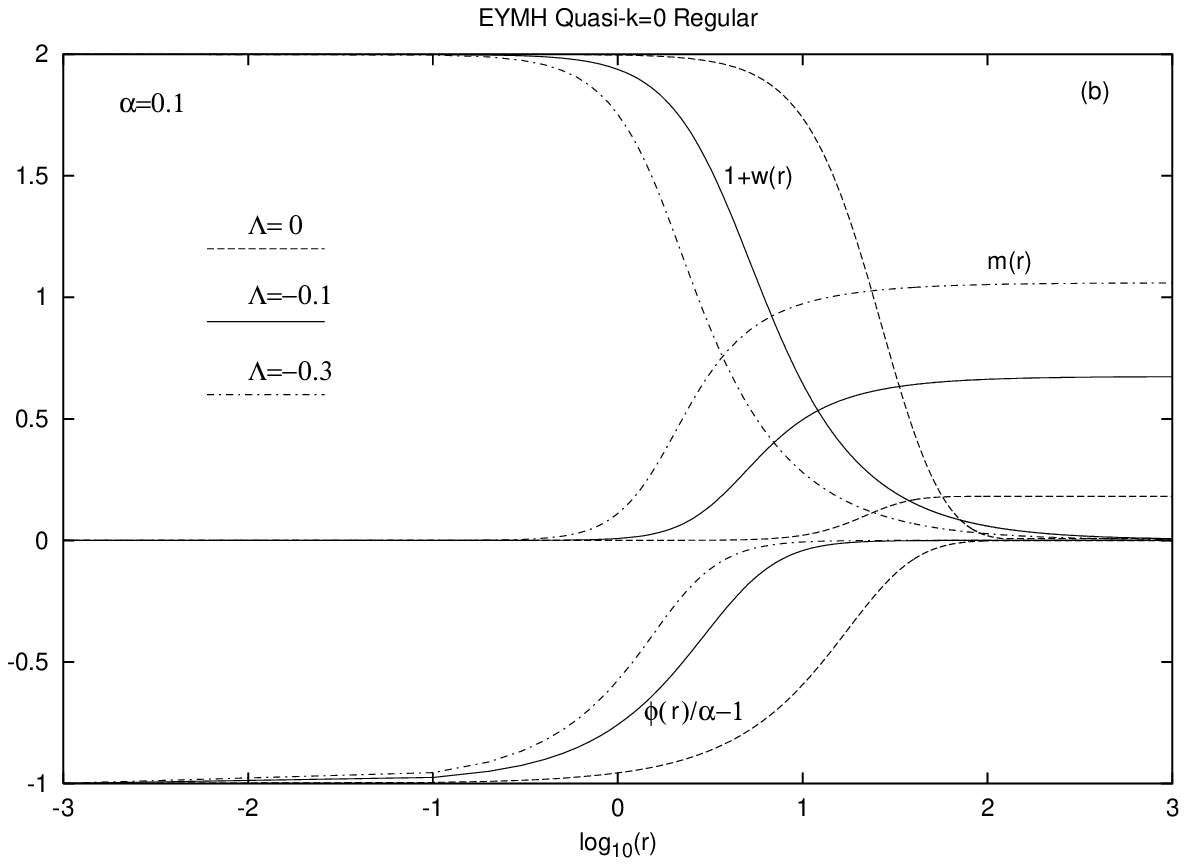,width=16cm}}
\end{picture}
\begin{center}
Figure 1b.
Quasi$-k=0$ regular $\alpha=0.1$; { } $\Lambda=0$, -0.1, -0.3\newline
\end{center}

\newpage
\begin{picture}(16,16)
\centering
\put(-2,0){\epsfig{file=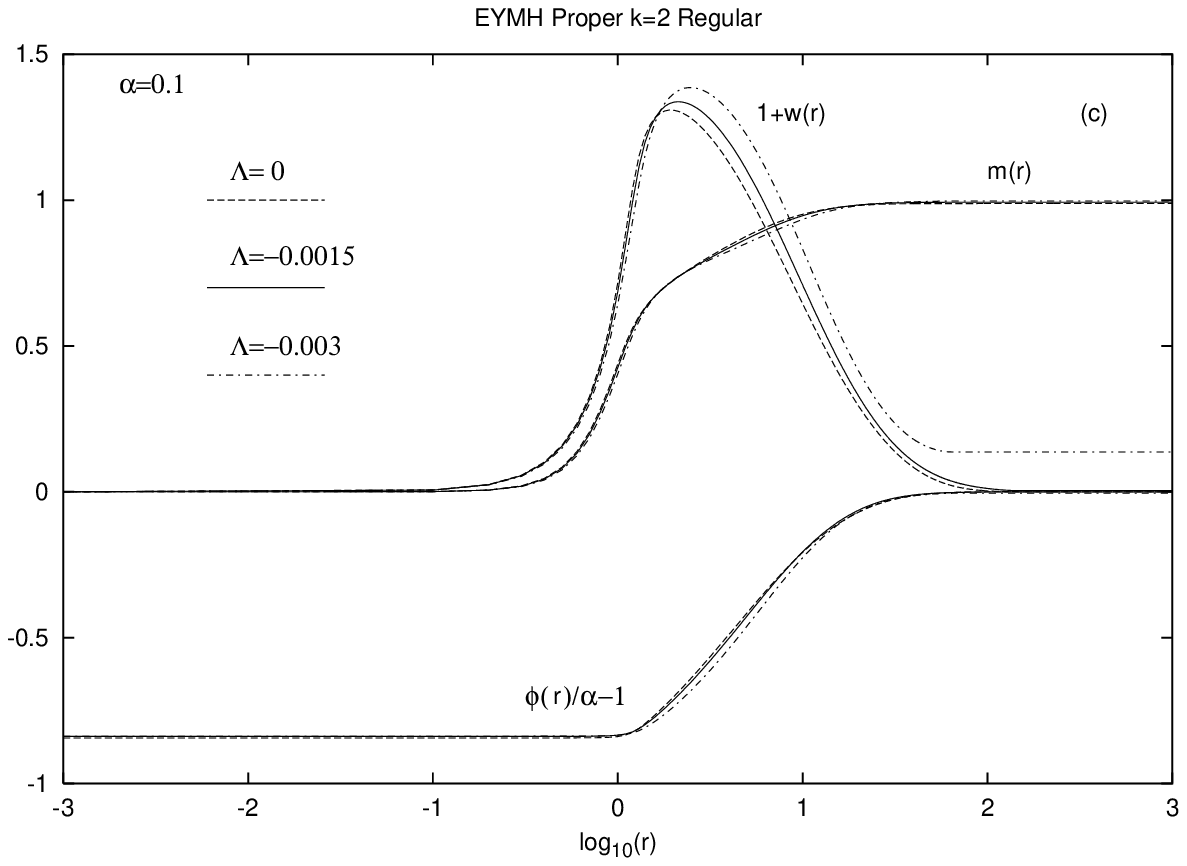,width=16cm}}
\end{picture}
\begin{center}
Figure 1c.
Proper $k=2$ regular $\alpha=0.1$; { } $\Lambda=0$, -0.0015, -0.003\newline
\end{center}

\newpage
\begin{picture}(16,16)
\centering
\put(-2,0){\epsfig{file=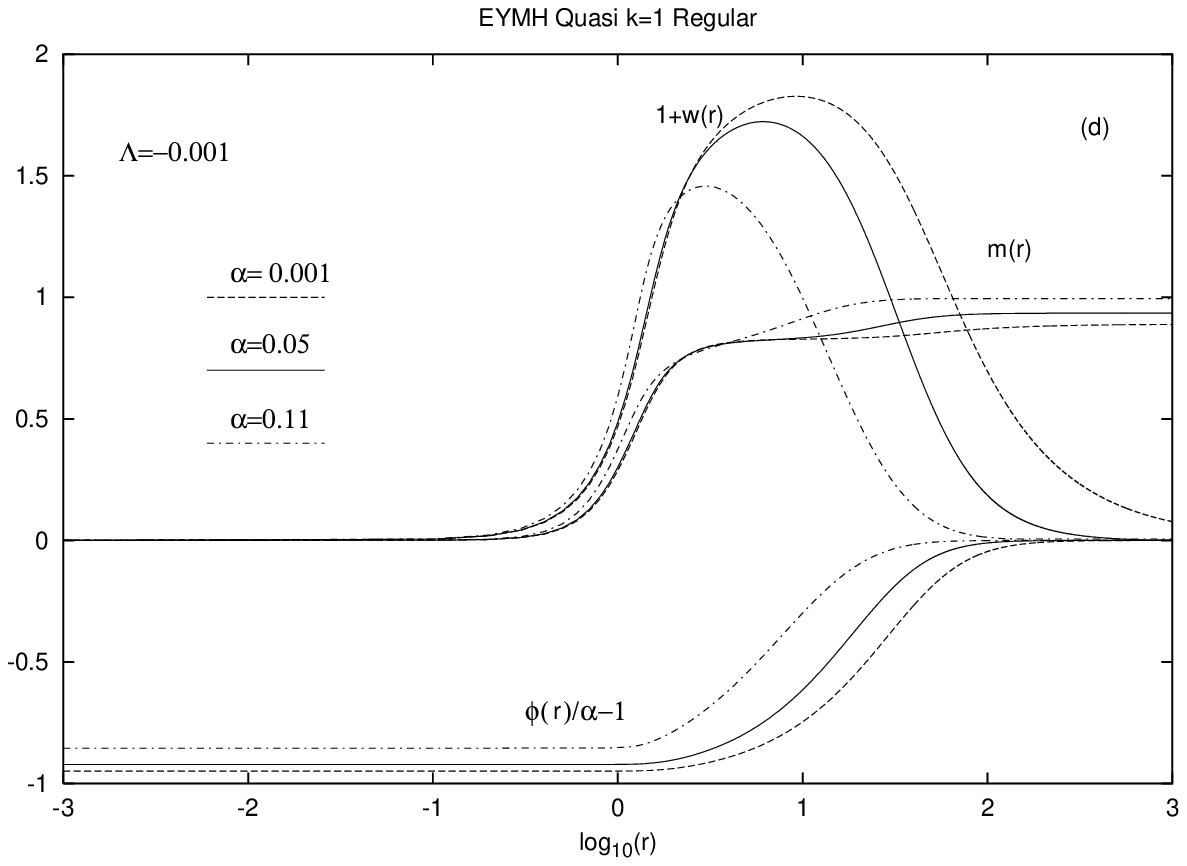,width=16cm}}
\end{picture}
\begin{center}
Figure 1d.
Quasi$-k=1$ Regular $\Lambda=-0.001$; { }$\alpha=0.001$, 0.005, 0.11\newline
\end{center}

\newpage
\begin{picture}(16,16)
\centering
\put(-2,0){\epsfig{file=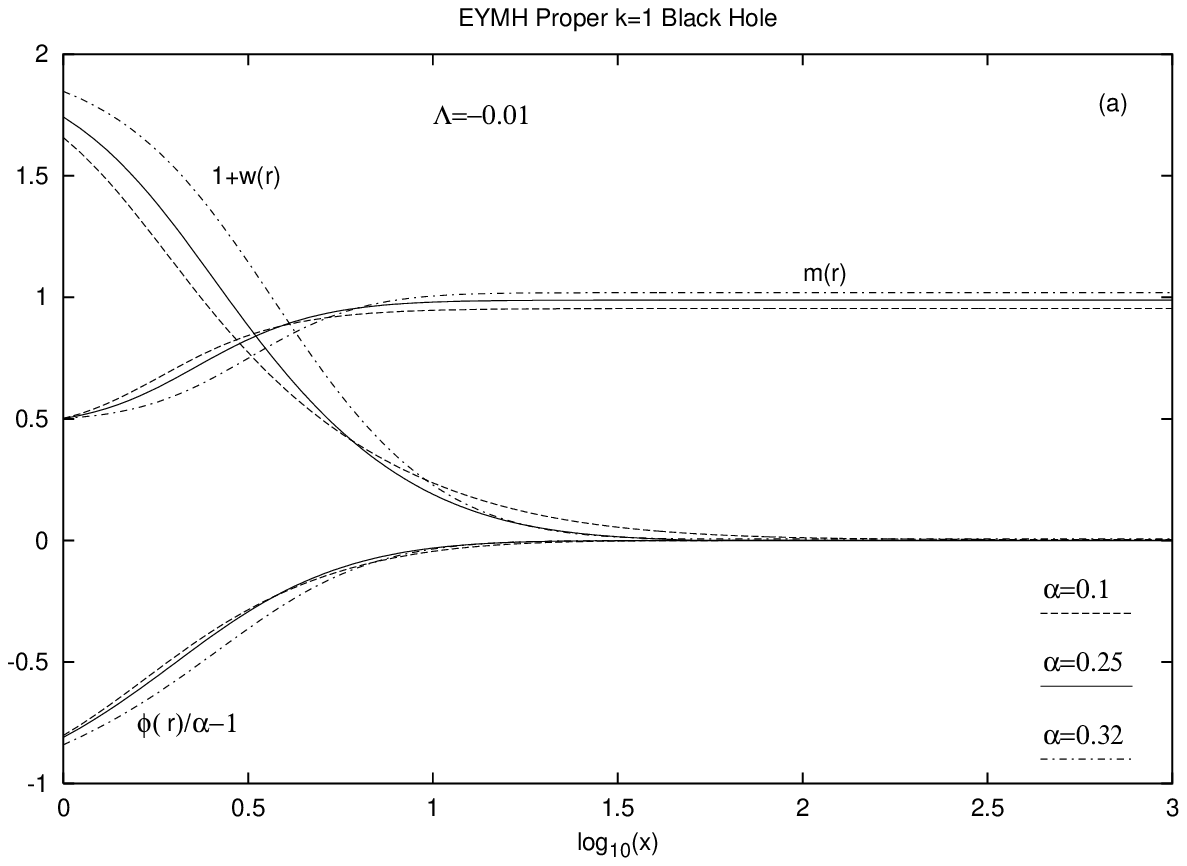,width=16cm}}
\end{picture}
\begin{center}
Figure 2a.
Proper $k=1$ black hole $\Lambda=-0.01$; { }$\alpha=0.1$, 0.25, 0.32\newline
\end{center}

\newpage
\begin{picture}(16,16)
\centering
\put(-2,0){\epsfig{file=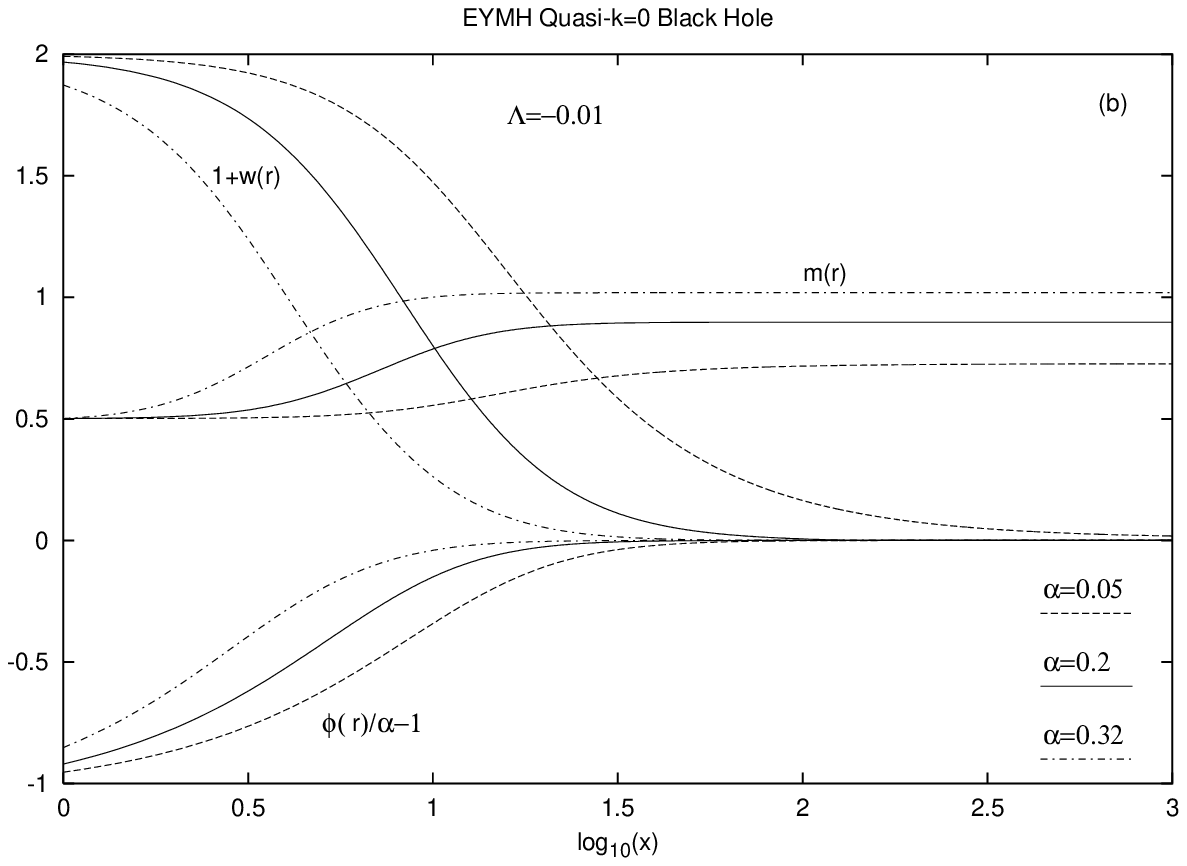,width=16cm}}
\end{picture}
\begin{center}
Figure 2b.
Quasi-$k=0$ black hole $\Lambda=-0.012$; { }$\alpha=0.05$, 0.2, 0.32\newline
\end{center}

\newpage
\begin{picture}(16,16)
\centering
\put(-2,0){\epsfig{file=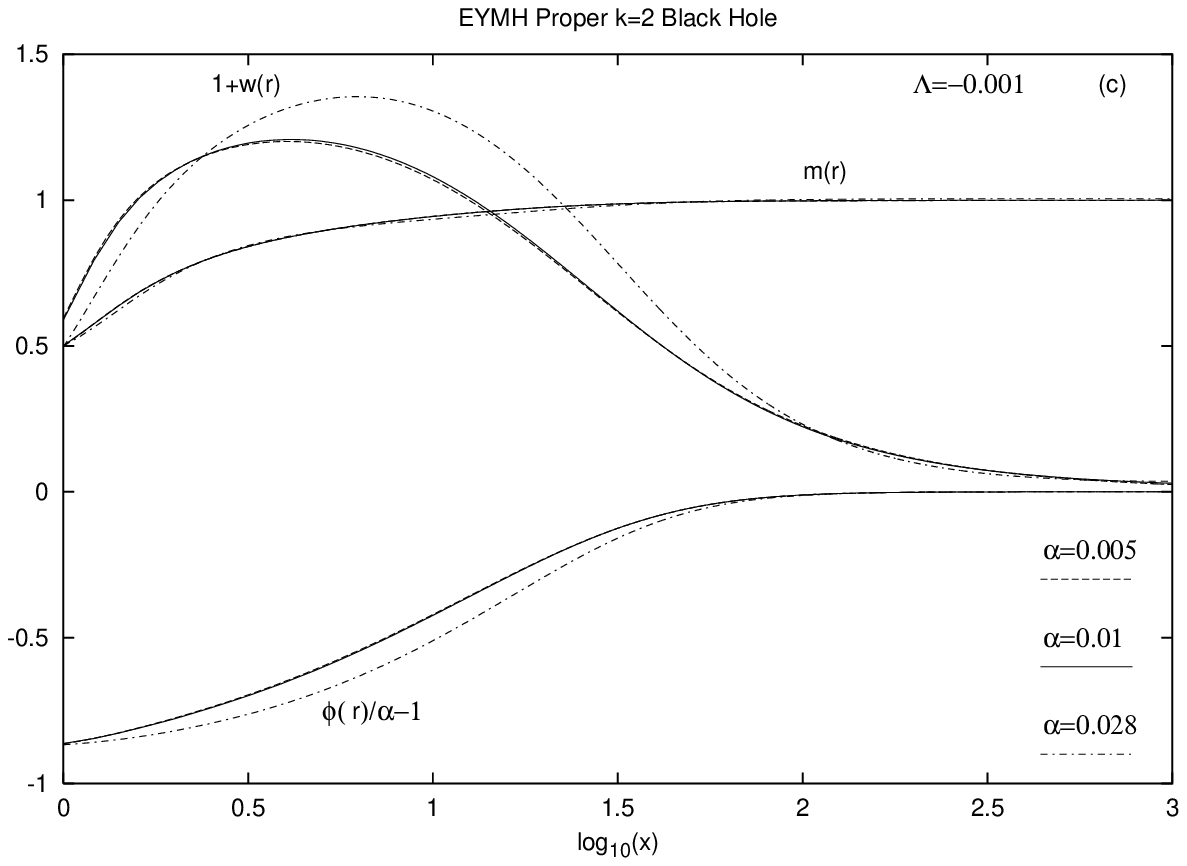,width=16cm}}
\end{picture}
\begin{center}
Figure 2c.
Proper $-k=2$ black hole $\Lambda=-0.001$; { }$\alpha=0.005$, 0.01, 0.028\newline
\end{center}

\newpage
\begin{picture}(16,16)
\centering
\put(-2,0){\epsfig{file=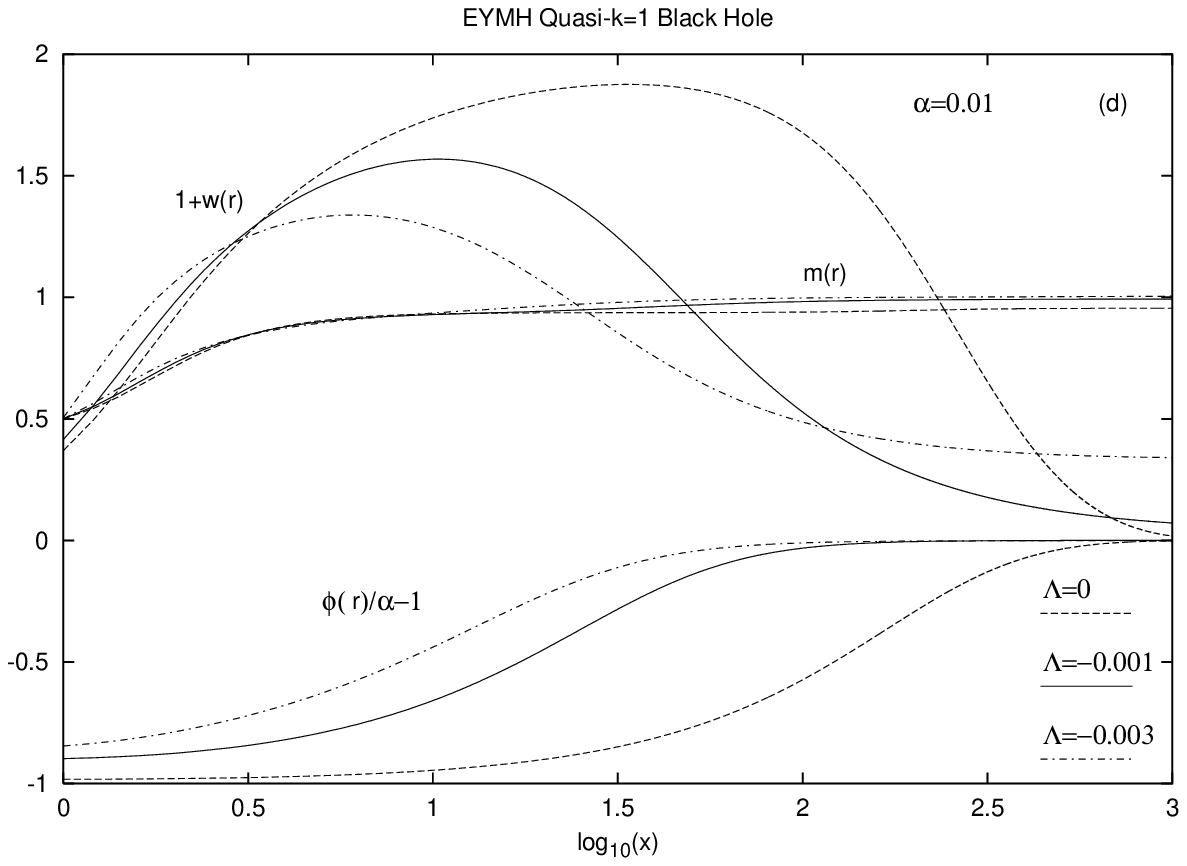,width=16cm}}
\end{picture}
\begin{center}
Figure 2d.
Quasi$-k=1$ black hole $\alpha=0.01$; { }$\Lambda=0.$, -0.001, -0.003\newline
\end{center}

\end{document}